\documentclass[12pt]{iopart}

\usepackage{amsfonts}

\def\be{\begin{equation}}
\def\ee{\end{equation}}

\def\bea{\begin{eqnarray}}
\def\eea{\end{eqnarray}}

\def\ben{\begin{enumerate}}
\def\een{\end{enumerate}}

\def\bea{\begin{eqnarray}}
\def\eea{\end{eqnarray}}

\begin{document}

\title[]{Algebraic Bethe Ans\"{a}tze and eigenvalue-based determinants for Dicke-Jaynes-Cummings-Gaudin quantum integrable models }
\author{Hugo Tschirhart and Alexandre Faribault}
\address{
Groupe de Physique Statistique, D\'epartement P2M, Institut Jean Lamour,\\ Universit\'e de Lorraine, CNRS (UMR 7198), 
Vand\oe uvre l\`es Nancy Cedex, F-54506, France
}
\ead{alexandre.faribault@univ-lorraine.fr}
\begin{abstract}

In this work, we construct an alternative formulation to the traditional Algebraic Bethe ansatz for quantum integrable models derived from a generalised rational Gaudin algebra realised in terms of a collection of spins 1/2 coupled to a single bosonic mode. The ensemble of resulting models which we call Dicke-Jaynes-Cummings-Gaudin models are particularly relevant for the description of light-matter interaction in the context of quantum optics. Having two distinct ways to write any eigenstate of these models we then combine them in order to write overlaps and form factors of local operators in terms of partition functions with domain wall boundary conditions. We also demonstrate that they can all be written in terms of determinants of matrices whose entries only depend on the eigenvalues of the conserved charges. Since these eigenvalues obey a much simpler set of quadratic Bethe equations, the resulting expressions could then offer important simplifications for the numerical treatment of these models. 
\end{abstract}

\date{\today}

\section{Introduction}

Finding their origin in the seminal works of Dicke  \cite{dicke} and Jaynes-Cummings \cite{JC} describing the interaction of two-level atoms with a single mode of the electromagnetic field, there has been major recent developments in the theory of quantum integrable models related to quantum optics. In this work we choose to generically call Dicke-Jaynes-Cummings-Gaudin (DJCG) this class of integrable models in which two-level systems are coupled to a single bosonic mode. 

The traditional approach \cite{jurco, gaudin,dukboson} to the DJCG-type integrable models has been to define them using a Holstein-Primakoff transformation on spin-models derived from the trigonometric  (XXZ) generalised Gaudin algebra \cite{gaudin,ortiz}. It allows one, in the large spin limit, to transform a spin degree of freedom into a bosonic mode. However, it was also later shown \cite{babelon,skrypnyk2008,skrypnyk2009} that one can directly build this class of integrable models through an appropriate realisation of the rational (XXX) generalised Gaudin algebra. This reinterpretation of such models highlights the strong relation the DJCG models have with the XXX Gaudin models \cite{gaudin} describing an ensemble of spins with long-range interaction. 

In light of this connection, it seemed reasonable to try to generalise a series of recent results obtained for spin-only models to the DJCG-family of integrable spin-boson models. Indeed, using a simple inversion $\hat{z} \to - \hat{z}$ of the quantization axis, Faribault and Schuricht \cite{faridet} introduced a determinant expression for the domain wall boundary partition function of rational Gaudin models \cite{gaudin,ortiz} realised in terms of $N$ spin-1/2 representations of the SU(2) algebra and showed how scalar products and certain form factors, can be written in the same fashion. One should know that the relation between scalar products and partition functions has also been explored in the context of spin chains in \cite{kostov}. The particular determinant form obtained in \cite{faridet} is explicitly written in terms of the eigenvalues of the conserved charges of the models and have ultimately allowed important numerical progress in the treatment of the central spin model's non-equilibrium dynamics, by allowing faster computation than traditional Slavnov-like determinant expressions. 

While they simply form a different realisation of the same algebra, models containing a bosonic mode differ fundamentally by no longer being bounded from above and below, since they now support an arbitrary number of excitations $M \in \mathbb{N}_0$. This slight difference prevented the straightforward generalisation of the constructions presented in \cite{faridet}, to spin-boson DJCG-models. 

In this work, such a generalisation is presented in a way which is as similar as can be to the approach used for spins-1/2. In the first section of the paper, a brief description of  the general Algebraic Bethe Anastz (ABA) for models derived from the rational generalized Gaudin algebra is given, irrespective of the realisation. We then proceed, in the next section, to a brief review of the eigenvalue-based determinant expressions for spin-only realisations. In section 4 we show how to build an alternative hole-like Bethe ansatz for spin-boson models and, in section 5, demonstrate how the domain wall boundary partition functions can be rewritten as eigenvalue-based determinants. The application of these two results to the calculation of various form factors is then presented in the last section of the paper. 

\section{Algebraic Bethe Ansatz for the rational generalised Gaudin algebra}
\label{abasec}

Let us first introduce the generalised Gaudin algebra defined by the operators $ \mathrm{S}^x(u), \mathrm{S}^y(u),\mathrm{S}^z(u)$ satisfying the commutation relations\cite{gaudin,ortiz}:

\bea
\left[\mathrm{S}^x(u),\mathrm{S}^y(v)\right] &=& i(Y(u,v) \mathrm{S}^z(u)- X(u,v)\mathrm{S}^z(v)),
\nonumber\\
\left[\mathrm{S}^y(u),\mathrm{S}^z(v)\right]  &=&i(Z(u,v) \mathrm{S}^x(u)- Y(u,v)\mathrm{S}^x(v)) ,
\nonumber\\
\left[\mathrm{S}^z(u),\mathrm{S}^x(v)\right] &=&i(X(u,v) \mathrm{S}^y(u)- Z(u,v)\mathrm{S}^y(v)) ,
\nonumber\\
\left[\mathrm{S}^\kappa(u),\mathrm{S}^\kappa(v)\right] &=& 0, \ \ \ \kappa=x,y,z,
\label{commut}
\eea

\noindent where $u,v \in \mathbb{C}$. Consistency  of the algebra imposed by the Jacobi identities can only be achieved when the functions $X,Y,Z$ fulfill the classical Yang-Baxter equation:

\bea
X(u, v)Y (v, w) + Y (w, u)Z(u, v) + Z(v, w)X(w, u) = 0.
\eea 

In this paper, we deal exclusively with one type of solutions to this equation, namely the rational (XXX) Gaudin algebra defined by

\bea
X(u,v)=Y(u,v) = Z(u,v) &=&  \frac{g}{u-v}\  , \ \  g \in \mathbb{R}.
\label{rational}
\eea

The operators

\bea
\mathrm{S}^2(u)\equiv\frac{1}{2}\left(\mathrm{S}^+(u)\mathrm{S}^-(u)+\mathrm{S}^-(u)\mathrm{S}^+(u)+2\mathrm{S}^z(u)\mathrm{S}^z(u)\right),
\label{tmatrix}
\eea

\noindent written here in terms of the raising and lowering Gaudin operators:

\bea
\mathrm{S}^\pm(u) \equiv \mathrm{S}^x(u) \pm i \mathrm{S}^y(u),
\eea

\noindent  are easily shown to all commute $\left[\mathrm{S}^2(u),\mathrm{S}^2(v)\right]=0$ for arbitrary complex parameters $u,v$. For a given number of excitations $M$, which is conserved for XXX models, the Quantum Inverse Scattering Method (QISM) \cite{korepin} and the resulting ABA allow one to find the eigenstates common to every $\mathrm{S}^2(u)$ by using the following generic construction:

\bea
\left|\lambda_1 \dots \lambda_M\right>\equiv \prod_{i=1}^M \mathrm{S}^+(\lambda_i) \left| 0 \right>.
\label{bethestate}
\eea

\noindent Here $\mathrm{S}^+(u)$ can be seen as an operator which creates a quasi-particle fully parametrized by a single complex variable $u \in \mathbb{C}$. The particle-pseudovacuum $\left|0 \right>$ is defined as a lowest weight vector for which $\mathrm{S}^-(u) \left|0 \right> = 0, $ $\forall \ u  \in \mathbb{C}$.

The action of the $\mathrm{S}^2(u)$ operator on a state of the form (\ref{bethestate}) can be obtained explicitly from the XXX commutation rules (\ref{commut}) as

\bea
\mathrm{S}^2(u) \left|\lambda_1 \dots \lambda_M\right> =   \left(\prod_{i=1}^M \mathrm{S}^+(\lambda_i)\right) \mathrm{S}^2(u) \left|0\right>  +  \left[\mathrm{S}^2(u) , \left(\prod_{i=1}^M \mathrm{S}^+(\lambda_i)\right)\right]\left|0\right>\nonumber\\
=E(\{\lambda\},u) \left|\lambda_1 ... \lambda_M\right> + \sum_{k=1}^M F_{k}(\{\lambda\},u) \left|\lambda_1 ...,\lambda_k\rightarrow u,... \lambda_M\right>,
\label{applic}
\eea

\noindent with 

\bea
E(\{\lambda\},u) &=&\ell(u)+ \sum_{i=1}^M \left(2F(u)X(u,\lambda_{i})+ \sum_{j\neq i}^MX(u,\lambda_{i})X(u,\lambda_{j})\right),
\label{eofu}
\eea

\noindent and

\bea
F_{k}(\{\lambda\},u) =2X(u,\lambda_{k})\left[F(\lambda_{k})+\sum_{j\neq k}X(\lambda_{j},\lambda_{k})\right],
\eea

\noindent provided the pseudovacuum  $\left|0 \right>$ is also an eigenstate of $\mathrm{S}^2(u)$ and $\mathrm{S}^z(u)$:

\bea
 \mathrm{S}^2(u)\left|0 \right> &=& \ell(u) \left|0 \right>
 \nonumber\\ \mathrm{S}^z(u)\left|0 \right> &=& F(u) \left|0 \right>.
\label{vacval}
 \eea

States of the form (\ref{bethestate}) therefore become common eigenstates of $\mathrm{S}^2(u) \ \forall \ u \in \mathbb{C}$ provided the $M$ rapidities $\lambda_i$ are solution of a set of coupled non-linear algebraic equations: the Bethe equations. For rational models (\ref{rational}), these equations are found by cancelling every $F_{k}(\{\lambda\},u)$ in eq. (\ref{applic}) and can be written, in general, as

\bea
F(\lambda_i)= \sum_{j=1 (\ne i)}^M \frac{g}{\lambda_i-\lambda_j},
\label{eq:RGeq}
\eea

\noindent with  $F(u)$ defined in (\ref{vacval}).

Numerically finding solutions to the Bethe equations as they are written above is a challenging task due in good part to the presence of cancelling divergences in the terms $\frac{1}{\lambda_i-\lambda_j}$ and $F(\lambda_i)$ when two (or more) rapidities $\lambda_i,\lambda_j$ approach the same value. While many efforts have been made over time to deal with these equations directly \cite{solving1,solving2,solving3,solving4}, the rewriting of the Bethe equations as an ensemble of $N$ quadratic equations \cite{babelon,baode1,baode2} has recently greatly simplified the numerical treatment of such systems \cite{bcsaraby,centralfaribault1,centralfaribault2,straeter}. 
It has indeed been shown that the Bethe equations for models with $F(u) = -\sum_{i=1}^N\frac{A_i}{\epsilon_i-u}+\frac{B}{2g} u + \frac{C}{2g}$ can be equivalently written as quadratic equations \cite{baode1,baode2}. A variety of XXX models do fall in that category including the spin-only realisations which are reviewed in the next section and the spin-boson realisations which are specifically studied in this work.  In the specific case of interest here ($A_i = \frac{1}{2}$ for spins-1/2) the quadratic Bethe equations can be written as:

\bea
\Lambda^2(\epsilon_j)+ \frac{B}{g} M - \frac{B \epsilon_i+C}{g}\Lambda(\epsilon_j) -\sum_{i\ne j}^N \frac{\Lambda(\epsilon_j)-\Lambda(\epsilon_i)}{\epsilon_j-\epsilon_i} = 0,
\eea

\noindent where we defined the $N$ variables $\Lambda(\epsilon_i) = \sum_{j=1}^M \frac{1}{\epsilon_i - \lambda_j}$. It should be pointed out that this approach is intimately linked to the Heine-Stietjes polynomial solutions to second-order Fuchsian equations which has also been exploited in certain works \cite{heinstil1,heinstil2,heinstil3,heinstil4,heinstil5}.

\section{Spin-models}

\subsection{Particle and hole-like Bethe Ans\"{a}tze}

Let us first briefly recall the constructions presented in \cite{faridet} for realisations of the generalised Gaudin algebra built exclusively out of $N$ local generators of a SU(2) spin algebra:

\bea
\mathrm{S}^+(u) = \sum_{i=1}^N \frac{S^+_i}{u-\epsilon_i}, \ \mathrm{S}^-(u) = \sum_{i=1}^N \frac{S^-_i}{u-\epsilon_i}, \ \mathrm{S}^z(u) = \frac{1}{g} - \sum_{i=1}^N \frac{S^z_i}{u-\epsilon_i},
\label{realspin}
\eea

\noindent with $g$ and $\{ \epsilon_1 \dots \epsilon_N\}$ arbitrary real parameters.

For these models, it is simple to understand that two alternative representations of any given eigenstate can be built: one using particle-like excitations through the repeated action of the raising operator on the particle-vacuum  $\prod_{i=1}^M \mathrm{S}^+(\lambda_i)\left|\downarrow \dots \downarrow\right>$ and the second one obtained using lowering operators acting on the hole-vacuum  $\prod_{i=1}^{N-M} \mathrm{S}^-(\mu_i)\left|\uparrow \dots \uparrow\right>$. We are here explicitly working with spins $S_i = \frac{1}{2}$, but a similar construction is possible for spins of arbitrary magnitude working with lowest/highest weights pseudovacuums \cite{faridet}. Since the spin algebra is symmetric in its highest/lowest weight configurations, both formulations of the ABA are formally identical, being simply related by a change of the quantization axis from $\hat{z} \to -\hat{z}$ which exchanges the roles of $\mathrm{S}^+(u)$ and  $\mathrm{S}^-(u)$ and replaces $\mathrm{S}^z(u)$ by $-\mathrm{S}^z(u)$ . The QISM is identically formulated in both cases since $\left|\uparrow \dots \uparrow\right>$ is a valid hole-pseudovacuum obeying all three properties:

\bea
\mathrm{S}^+(u) \left|\uparrow \dots \uparrow\right> = 0, \ \
\mathrm{S}^2(u) \left|\uparrow \dots \uparrow\right> =  \ell_{\uparrow}(u)\left|\uparrow \dots \uparrow\right>,
\nonumber\\
-\mathrm{S}^z(u) \left|\uparrow \dots \uparrow\right> =  F_{\uparrow}(u) \left|\uparrow \dots \uparrow\right>.
\eea

One then finds two sets of Bethe equations which, for $M$ particles and $(N-M)$ holes, read:

\bea
-\frac{1}{2}\sum_{k=1}^N \frac{1}{\epsilon_k-\lambda_j} + \frac{1}{g}&=&  \sum_{k \ne j}^{M} \frac{1}{\lambda_j-\lambda_k} \ \ \ \mathrm{(particles)}
\nonumber\\
-\frac{1}{2}\sum_{k=1}^N \frac{1}{\epsilon_k-\mu_j} - \frac{1}{g}&=& \sum_{k \ne j}^{N-M} \frac{1}{\mu_j-\mu_k} \ \ \  \mathrm{(holes)}
\label{doublebethe}
\eea

\noindent and whose eigenvalue-based quadratic equations are respectively given by:

\bea
\left[\Lambda^\lambda(\epsilon_i)\right]^2&=&  \sum_{j \ne i}^N \frac{\Lambda^\lambda(\epsilon_i)-\Lambda^\lambda(\epsilon_j)}{\epsilon_i-\epsilon_j} + \frac{2}{g} \Lambda^\lambda(\epsilon_i) \ \ \ \mathrm{(particles)}
\nonumber\\
\left[\Lambda^\mu(\epsilon_i)\right]^2&=&  \sum_{j \ne i}^N \frac{\Lambda^\mu(\epsilon_i)-\Lambda^\mu(\epsilon_j)}{\epsilon_i-\epsilon_j} - \frac{2}{g} \Lambda^\mu(\epsilon_i). \ \ \ \mathrm{(holes)}
\label{doublebethequad}
\eea

Solutions to these two sets of equations both define the same ensemble of $M$ particles eigenstates of the transfer matrix $\mathrm{S}^2(u)$. The explicit expression of $\mathrm{S}^2(u)$, for the realisation $ (\ref{realspin})$ has a series of $N$ poles at $u=\epsilon_i$ whose residues correspond to the $N$ commuting conserved operators:

\bea
R_i = -\frac{ 2S^z_i }{g}  + \sum_{j=1(\ne i)}^N\frac{2\vec{S}_i \cdot \vec{S}_j }{\epsilon_i-\epsilon_j}, 
\label{conservedspin}
\eea

\noindent whose eigenvalues $r_i$ are read off the residues of the $\mathrm{S}^2(u)$ eigenvalues (\ref{eofu}). In both representations (for spins-$\frac{1}{2}$ only) these eigenvalues read:

\bea
r^\lambda_i = -\sum_{j=1}^{M}\frac{1}{\epsilon_i - \lambda_j}+ \frac{1}{2}\sum_{j=1 (\ne i)}^N \frac{1}{\epsilon_i-\epsilon_j} + \frac{1}{g}
\nonumber\\
 r^\mu_i = -\sum_{j=1}^{N-M}\frac{1}{\epsilon_i - \mu_j} +  \frac{1}{2}\sum_{j=1 (\ne i)}^N \frac{1}{\epsilon_i-\epsilon_j} - \frac{1}{g}
\label{eigenspin}
\eea

\noindent which explicitly depend on the state only through $\Lambda^\lambda(\epsilon_i) = \sum_{j=1}^M \frac{1}{\epsilon_i - \lambda_j}$ and $\Lambda^\mu(\epsilon_i) = \sum_{j=1}^{N-M} \frac{1}{\epsilon_i - \mu_j}$, hence the name "eigenvalue-based" used here to describe expressions built out of the $\Lambda(\epsilon_i)$ variables. One should also know that, since the parameters  $\lambda$ and $\mu$ have to be either real or come in complex conjugate pairs, the resulting  $\Lambda(\epsilon_i)$ are systematically real for any eigenstate of the system.

The correspondence between the two representations of a given eigenstate is then easily found by picking the solutions which give the same ensemble of eigenvalues $r_k$. Doing so, directly shows that the transformation:

\bea
\Lambda^\mu(\epsilon_i) = \Lambda^\lambda(\epsilon_i) - \frac{2}{g}
\label{reptransfo}
\eea
\noindent allows one to simply go from one representation to the other for an arbitrary eigenstate, as defined by any solution to either form of the Bethe equations (\ref{doublebethe}) or, alternatively (\ref{doublebethequad}). 

\subsection{Partition functions, normalisation and form factors}
\label{spindets}
One can then compute the scalar product of a Bethe state (defined by (\ref{bethestate}) using an arbitrary set $\{\lambda_1 \dots \lambda_M\}$)  with any eigenstate common to $S^z_1 \dots S^z_N$ containing $M$ up-pointing spins labelled by the indices $\{i_1\dots i_M\}$; the $N-M$ other spins $\{\overline{i}_1\dots \overline{i}_{N-M}\}$ therefore pointing down. We can alternatively write this state on which we project as either a particle or a hole-like construction $\left| \uparrow_{\{i_1 \dots i_{M}\}} \right>=\left| \downarrow_{\{\overline{i}_1 \dots \overline{i}_{N-M}\}} \right>$. The scalar product of the Bethe state with this particular basis state was shown \cite{faridet} to be given by the determinant of an $M\times M$ matrix:

\bea
\left< \uparrow_{\{i_1 \dots i_{M}\}} \right.\left| \lambda_1 \dots \lambda_M \right> = \left< \uparrow_{\{i_1 \dots i_{M}\}} \right| \prod_{i=1}^M \mathrm{S}^+(\lambda_i) \left| \downarrow \dots \downarrow\right> 
&=& \mathrm{Det} J_{M\times M} 
\eea
\bea
J_{ab}  &=&  \left\{ \begin{array}{cc}
\displaystyle\sum_{c =1(\ne a)}^M  \frac{1}{\epsilon_{i_a} -\epsilon_{i_c} }-\Lambda^\lambda(\epsilon_{i_a}) & a= b
\\
\frac{1}{\epsilon_{i_a} -\epsilon_{i_b} }
& a\ne b \end{array}\right. 
\label{overlaps}
\eea
\noindent by using the fact that it is a rational function of the $\lambda_i$ parameters which obeys a simple recursion relation. An identical construction is possible for hole-like states $\left< \uparrow_{\{i_1 \dots i_{M}\}} \right.\left| \mu_1 \dots \mu_{N-M} \right> = \left< \downarrow_{\{\overline{i}_1 \dots \overline{i}_{N-M}\}} \right|\prod_{i=1}^{N-M}\mathrm{S}^-(\mu_i) \left| \uparrow \dots \uparrow\right>$, which, by symmetry, are given by the same form of determinant, this time of an $(N-M)\times(N-M)$ matrix. That matrix is then defined by the $N-M$ values of $\epsilon$ associated with the spins which are pointing down in the bra and the replacement $\Lambda^\lambda \to \Lambda^\mu $.

Having access to the two representations of any given eigenstate allows one to rewrite their scalar product (even for two eigenstates found at different values of $g$) as partition functions with domain wall boundary conditions :

\bea
\left<\mu_1 \dots \mu_{N-M}\right.|\left.\lambda_1 \dots \lambda_M\right>&=& \left< \uparrow \dots \uparrow \right| \left(\prod_{i=1}^{N-M} \mathrm{S}^+(\mu_i) \right) \left(\prod_{i=1}^{M} \mathrm{S}^+(\lambda_i)\right)\left| \downarrow \dots \downarrow \right> \nonumber\\ &\equiv& \left< \uparrow \dots \uparrow \right| \left(\prod_{i=1}^{N} \mathrm{S}^+(\nu_i) \right)\left| \downarrow \dots \downarrow \right>,
\eea

\noindent where $\{\nu_1 \dots \nu_N\}  = \{\lambda_1 \dots \lambda_M\} \cup \{\mu_1 \dots \mu_{N-M}\} $. They are therefore writable, for an arbitrary ensemble $\{\nu_1 \dots \nu_N\}$, as the determinant of the $N\times N$ version of the matrix defined above:

\bea
 \left< \uparrow \dots \uparrow \right| \left(\prod_{i=1}^{N} \mathrm{S}^+(\nu_i) \right)\left| \downarrow \dots \downarrow \right> =\mathrm{Det} J_{N\times N}  \ \ \ \forall \ \{\nu_1 \dots \nu_N\} \in \mathbb{C}^N 
\eea

\noindent which is explicitly written in terms of the variables $\Lambda^{\nu}(\epsilon_i) = \sum_{j=1}^N \frac{1}{\epsilon_i - \nu_j} = \Lambda^{\lambda}(\epsilon_i)+\Lambda^{\mu}(\epsilon_i)$:

\bea
J_{ab} &=&
\left\{ \begin{array}{cc}
\displaystyle\sum_{c =1(\ne a)}^N  \frac{1}{\epsilon_{a} -\epsilon_{c} }-\Lambda^\lambda(\epsilon_a) -\Lambda^{\mu}(\epsilon_a) & a= b
\\
\frac{1}{\epsilon_{a} -\epsilon_{b} }
& a\ne b \end{array}\right. .
\label{overlapsspin}
\eea

While this expression is valid for the overlaps of arbitrary particle-like and hole-like states, one should keep in mind that a generic particle-state (built out of arbitrary $\{\lambda_1 \dots \lambda_M\}$) cannot necessarily be rewritten as an equivalent hole-like representation. However, for eigenstates of $\mathrm{S^2(u)}$ (defined by a solution to the Bethe equations) we showed that such a hole-representation not only exists, but is also quite simple to find using eq. (\ref{reptransfo}). 

Consequently, for a given eigenstate, its particle representation ($\left|\lambda_1 \dots \lambda_M\right>$) and hole representation ($\left|\mu_1 \dots \mu_{N-M}\right>$) correspond to the same normalised state $\left|\lambda_1 \dots \lambda_M\right>_n$ and only differ by a constant prefactor: $\left|\lambda_1 \dots \lambda_M\right> = N_\lambda\left|\lambda_1 \dots \lambda_M\right>_n$ and $\left|\mu_1 \dots \mu_{N-M}\right>=N_\mu\left|\lambda_1 \dots \lambda_M\right>_n$.  The scalar product between both representations of the SAME state and the projections on an arbitrary eigenstate of $\{S^z_1 \dots S^z_N\}$ gives us access to the individual normalisation of both representations \cite{centralfaribault1} since they respectively give us the product and ratio of these constants:

\bea
\left< \{ \mu \}\  \right| \left.\{ \lambda \} \right> = \mathrm{Det}K  =   N_\lambda N_\mu 
\nonumber\\
\frac{\left< \uparrow_{\{i_1 \dots i_{M}\}}  \right.\Big| \{ \lambda \} \Big>}{ {\left< \uparrow_{\{i_1 \dots i_{M}\}}\right. \Big| \{ \mu \} \Big> }}
= \frac{\mathrm{Det}J^\lambda_{M \times M}}{\mathrm{Det}J^\mu_{(N-M) \times (N-M)}} = \frac{N_\lambda}{N_\mu}.
\eea

\noindent These two relations are evidently sufficient to compute the squared norm of both representations allowing one to properly normalize the states. While the determinant expressions show that both $N_\mu$ and $N_\lambda$ have to be real, they can still in principle differ by a sign corresponding to a $\pi$ phase between both representations. The sign of the product $N_\lambda N_\mu$ allows one to simply detect this $\pi$ phase for any given eigenstate and therefore correct this possible phase shift between both representations of the normalised eigenstate.

Form factors for the various local raising and lowering operators  $S^\pm_i$ are also easily writable as a similar determinant. Since these form factors can only be non-zero when they involve two states whose number of excitations differs by one and since the local spin-raising operators are simply the residues of the Gaudin operator $\mathrm{S}^+(u)$ (\ref{realspin}), one simply needs to take the appropriate limit of the previous determinant to find: 

\bea
&&\left< \{ \mu_1 \dots \mu_{N-M} \} \right| S^+_k \left| \{ \lambda_1 \dots \lambda_{M-1} \} \right> \nonumber\\&&= \lim_{u \to \epsilon_k} (u-\epsilon_k)\left< \{ \mu_1 \dots \mu_{N-M} \} \right| \mathrm{S}^+(u) \left| \{ \lambda_1 \dots \lambda_{M-1} \} \right> \nonumber\\&&= \lim_{u \to \epsilon_k} (u-\epsilon_k) \mathrm{Det} J(\{\mu_1 \dots \mu_{N-M},u, \lambda_1 \dots \lambda_{M-1}\}) = \mathrm{Det} J^{\hat{k}},
\eea

\noindent where $J^{\hat{k}}$ is the $(N-1)\times (N-1)$ matrix equivalent to (\ref{overlapsspin}) from which line and column $k$ have been removed while, at the same time, the sums in the diagonal elements now exclude $\epsilon_k$:

\bea
J^{\hat{k}}_{ab} &=&
 \left\{ \begin{array}{cc}
\displaystyle\sum_{c =1(\ne a,k)}^N  \frac{1}{\epsilon_{a} -\epsilon_{c} }-\Lambda^\lambda(\epsilon_a) -\Lambda^{\mu}(\epsilon_a) & a= b
\\
\frac{1}{\epsilon_{a} -\epsilon_{b} }
& a\ne b \end{array}\right.  \ \mathrm{ with} \ (a,b \ne k).
\label{ffspinraisingforspins}
\eea

While no similar eigenvalue-based representation of the $S^z_i$ form factors has been proposed yet, they will be built explicitly for DJCG-models in section \ref{numbersection}. This result will then be straightforwardly generalisable to spin-only realisations as we will mention.

An important aspect of these expressions is that they allow calculations of scalar products and form factors using exclusively  the variables $\Lambda(\epsilon_i)$ which are much simpler to solve for since they obey quadratic Bethe equations. Therefore, they have been instrumental in allowing the fast and efficient numerical calculations necessary to study the fully quantum non-equilibrium dynamics of the central spin model \cite{centralfaribault1,centralfaribault2} for example. While the determinants are those of larger matrices ($N\times N$) than Slavnov's $M\times M$ determinants \cite{slavnov,slavnovlinks}, they  can still provide  more efficient numerics. Indeed, being expressed in terms of the $\{\Lambda^\lambda(\epsilon_1) \dots \Lambda^\lambda(\epsilon_N) \}$ variables, their use avoids the complicated extraction of the $\{\lambda_1 \dots \lambda_M\}$ variables corresponding to a given $\{\Lambda^\lambda(\epsilon_1) \dots \Lambda^\lambda(\epsilon_N) \} $. Generalising these constructions to other realisations of the Gaudin algebra (\ref{commut}) can, for the same reason, become a particularly useful exercise.

\section{Spin-boson realisations}
\label{spinboson}

In generalising the previous results to models which include a bosonic mode, we explicitly restrict ourselves to the particular realisation which combines a single bosonic degree of freedom and $N$ spins-1/2 and from which the DJCG-models can be derived. It explicitly reads:

\bea
\mathrm{S}^+(u) = b^\dag + \sum_{j=1}^N \frac{V}{u-\epsilon_j} S^+_j \ ,\  \mathrm{S}^-(u) = b + \sum_{j=1}^N \frac{V}{u-\epsilon_j} S^-_j  ,\nonumber\\  \mathrm{S}^z(u) = \frac{\omega-u}{2V} - \sum_{j=1}^N \frac{V}{u-\epsilon_j} S^z_j,
\label{sbreal}
\eea

\noindent with $b^\dag,b$ obeying canonical bosonic commutation rules. The conserved charges, found by looking at the residues of $\mathrm{S}^2(u)$, are then given by \cite{dukboson,skrypnyk2009}:

\bea
R_i =  \left(\epsilon_i - \omega\right) S^z_i+V\left(b^\dag S^-_i+bS^+_i\right) + \sum_{j\ne i}^N \frac{2V^2}{\epsilon_i-\epsilon_j} \vec{S}_i\cdot \vec{S}_j,
\label{consboson}
\eea

\noindent in terms of which, the full generating function is given by

\bea
\mathrm{S}^2(u) = \sum_{i=1}^N \frac{R_i}{u-\epsilon_i} + \left[b^\dag b + \sum_{i=1}^N S^z_i \right] + \frac{1}{2} + \left(\frac{\omega-u}{2V}\right)^2+\frac{3}{4} \sum_{i=1}^N \frac{V^2}{(u-\epsilon_i)^2},
\nonumber\\
\label{generating}
\eea

\noindent where the $\frac{3}{4}$ factors are simply the Casimir invariant of each local spin-$\frac{1}{2}$. 

A typical application of the QISM on this system \cite{ortiz} would be carried out by using the particle-like construction creating excitations above a pseudo-vacuum defined by a fully down-polarised spin sector and an empty bosonic mode. It leads to eigenstates of the form:

\bea
\left|\lambda_1 \dots \lambda_M\right> = \prod_{i=1}^M \mathrm{S}^+(\lambda_i) \left|0; \downarrow, \dots \downarrow\right>.
\label{bethelikeboson}
\eea

This vacuum $\left|0; \downarrow, \dots \downarrow\right>$ is indeed an eigenstate of both $\mathrm{S}^2(u)$ and $\mathrm{S}^z(u)$ and is annihilated by any $\mathrm{S}^-(u)$. Thus, it allows the direct use of the construction described in section \ref{abasec}.  The resulting ABA shows that the previous state becomes an eigenstate of $\mathrm{S}^2(u)$ whenever $\{\lambda_1 \dots \lambda_M\}$ is a solution of the algebraic system of Bethe equations:

\bea
\frac{\omega-\lambda_i}{2V^2} + \frac{1}{2} \sum_{k=1}^N \frac{1}{\lambda_i - \epsilon_k} = \sum_{j\ne i}^M \frac{1}{\lambda_i - \lambda_j}.
\label{bethebosrap}
\eea

The eigenvalues of the conserved charges (\ref{consboson}) are then respectively given by:

\bea
r_i = \frac{V^2}{2} \left(\sum_{j\ne i}^N \frac{1}{\epsilon_i-\epsilon_j}\right) -\frac{\epsilon_i-\omega}{2} - V^2 \sum_{j=1}^M \frac{1}{\epsilon_i - \lambda_j}.
\label{ripart}
\eea 

Since they have a fixed number $M$ of particle excitations these states are not only eigenstates of every $R_i$ integrals but are, as well, eigenstates of the "number" operator $b^\dag b + \sum_{i=1}^N S^z_i $  with eigenvalue $M-\frac{N}{2}$. According to (\ref{generating}) they are therefore eigenstates of $\mathrm{S}^2(u)\  \forall \ u \in \mathbb{C}$ with eigenvalue $\sum_{i=1}^N \frac{r_i}{u-\epsilon_i} + M-\frac{N}{2} +\frac{1}{2} + \left(\frac{\omega-u}{2V}\right)^2+\frac{3}{4} \sum_{i=1}^N \frac{V^2}{(u-\epsilon_i)^2} $.

This realisation allows an equivalent description in terms of eigenvalue-based quadratic Bethe equations \cite{baode1} given, this time, by:

\bea
 \left[ \Lambda(\epsilon_i)\right]^2 = \sum_{j \ne i}^N \frac{ \Lambda(\epsilon_i)- \Lambda(\epsilon_j)}{\epsilon_i-\epsilon_j}  -\frac{\epsilon_i - \omega}{V^2} \Lambda(\epsilon_i)+ \frac{M}{V^2},
\label{quadbos}
\eea

\noindent with $\Lambda(\epsilon_i) =  \displaystyle \sum_{j=1}^M \frac{1}{\epsilon_i - \lambda_j}$ still.

 Central to the constructions presented before for spin-only models is the inherent symmetry between maximal and minimal weight states which allowed one to define an equivalent hole-like Bethe ansatz. However, for these spin-boson DJCG-models, the unboundedness of the bosonic part gives rise to a problem in defining the alternative hole-pseudovacuum. Still, simple considerations about the total number of quasi-particles $M$, which is a conserved quantity in XXX (and XXZ) models, allow one to suppose that a  possible hole-vacuum state could be chosen as $\left| M; \uparrow \uparrow ... \uparrow \right>$. Indeed, acting on this state with $N$ lowering operators $\prod_{i=1}^N \mathrm{S}^-(\mu_i)\left| M; \uparrow\uparrow   \dots \uparrow \right>$ produces a state which spans the same subspace of the full Hilbert space as $\prod_{i=1}^M \mathrm{S}^+(\lambda_i)\left| 0; \downarrow \downarrow  \dots\downarrow \right>$. Both states are indeed linear superpositions of tensor-product states containing any number of bosonic excitations $ N_b \in \left[0,M\right]$ and any subset of $M-N_b$ spins (picked out of the $N$ available ones) pointing up. However, it remains to be shown that $\prod_{i=1}^N \mathrm{S}^-(\mu_i)\left| M; \uparrow \uparrow  \dots \uparrow\right>$ can provide an alternative representation of the eigenstates of the system.

Working with a typical QISM construction for these models, one would explicitly calculate $\mathrm{S}^2(u) \left[\prod_{i=1}^N \mathrm{S}^-(\mu_i)\right]\left| M; \uparrow \uparrow  \dots \uparrow \right> = \left[\prod_{i=1}^N \mathrm{S}^-(\mu_i) \right] \mathrm{S}^2(u) \left| M; \uparrow \uparrow  \dots \uparrow \right> + [ \mathrm{S}^2(u) , \prod_{i=1}^N  \mathrm{S}^-(\mu_i)]\left| M; \uparrow \uparrow \dots \uparrow \right>$ and look for the set of equations for $\{\mu_1  \dots \mu_N\}$ whose solutions reduce to $ \mathrm{S}^2(u) \left[\prod_{i=1}^N  \mathrm{S}^-(\mu_i)\right]\left| M; \uparrow \uparrow  \dots \uparrow \right> = E(u)\left[\prod_{i=1}^N  \mathrm{S}^-(\mu_i) \right]  \left| M; \uparrow \uparrow \dots \uparrow \right>  $. By being an eigenstate of  $\mathrm{S}^2(u)$, a proper pseudo-vacuum would make the term $\left[\prod_{i=1}^N \mathrm{S}^-(\mu_i) \right] \mathrm{S}^2(u) \left| M; \uparrow \uparrow  \dots \uparrow \right> $ trivially proportional to the original ansatz $\left[\prod_{i=1}^N \mathrm{S}^-(\mu_i)\right]\left| M; \uparrow \uparrow  \dots \uparrow \right>$. However, in the case at hand here, this potential "hole-vacuum" is NOT an eigenstate of $\mathrm{S}^2(u)$:

\bea
 \mathrm{S}^2(u) \left| M;\uparrow  \dots \uparrow \right> &=&  \left[ \mathrm{S}^2_z(u)+\frac{1}{2} \mathrm{S}^+(u) \mathrm{S}^-(u)+\frac{1}{2} \mathrm{S}^-(u) \mathrm{S}^+(u)\right]  \left| M;  \uparrow  \dots \uparrow \right> 
\nonumber\\
&=&
\left[ \mathrm{S}^2_z(u)+\frac{1}{2}\left[ \mathrm{S}^+(u), \mathrm{S}^-(u)\right]  + \mathrm{S}^-(u) \mathrm{S}^+(u)\right]  \left| M; \uparrow  \dots \uparrow\right> 
\nonumber\\
&=&
\ell(u)  \left| M;   \uparrow  \dots \uparrow \right>+\sqrt{M+1} \ \mathrm{S}^-(u)\left| M+1; \uparrow  \dots \uparrow \right>.\nonumber\\
\eea

\noindent since $\left| M;\uparrow  \dots \uparrow\right>$ is an eigenstate of $ \mathrm{S}^2_z(u)+\frac{1}{2}\left[ \mathrm{S}^+(u), \mathrm{S}^-(u)\right] $ whose eigenvalue we write $\ell(u)$. The second term can be written in that way since $\mathrm{S}^+(u)$ only affects the bosonic part because every spin in the vacuum state is already pointing up. The fact that we do not have here a "proper" hole-vacuum turns this reversed problem into one which could be approached using Sklyanin's separation of variables \cite{sklyanin} and the resulting Functional Bethe Ansatz (FBA) \cite{amico}. We show however that, although it is slighty more complex, one can still follow closely the QISM approach to formulate an alternative algebraic Bethe anastz. 

The standard application of the generating function $\mathrm{S}^2(u)$ on the hole-like ansatz $|\mu\rangle\equiv\prod_{i=1}^N  \mathrm{S}^-(\mu_i)\left| M; \uparrow \dots \uparrow \right>$ would result in the generic form:

\bea
\mathrm{S}^2(u) &=& \tilde{E}(u,\{\mu_1 ... \mu_n\}) |\mu\rangle + \sum_{k=1}^N F_k(u,\{\mu_1 ... \mu_n\})
|\mu\rangle_{\mu_k \to u}+ \nonumber\\ &&\sqrt{M+1}\left(\prod_{i=1}^N  \mathrm{S}^-(\mu_i) \right)\mathrm{S}^-(u) \left| M+1; \uparrow \dots \uparrow \right>.\eea

It turns out, and that is easily verified numerically for small systems, that the last two terms one might naively consider to be the unwanted terms, cannot be cancelled out for arbitrary $u \in \mathbb{C}$. When used in conjunction with a proper pseudo-vacuum, the QISM does lead to a simple structure where both the correct eigenvalue and the unwanted terms naturally appear. Here, due the "faulty" vacuum in use, $ \tilde{E}(u,\{\mu_1 ... \mu_n\})$ is not the actual eigenvalue of $\mathrm{S}^2(u)$. The unwanted terms which can be cancelled must contain a $u$-dependent piece $\Delta E(u,\{\mu_1 ... \mu_n\}) |\mu\rangle $ taken from the term $\tilde{E}(u,\{\mu_1 ... \mu_n\}) |\mu\rangle $ and the actual eigenvalue then become $\tilde{E}(u,\{\mu_1 ... \mu_n\})-\Delta E(u,\{\mu_1 ... \mu_n\}) $. Having no a priori knowledge of the function  $\Delta E(u,\{\mu_1 ... \mu_n\})$ one should use for a given $u$ in order to build the correct unwanted term: $\Delta E(u,\{\mu_1 ... \mu_n\})  |\mu\rangle + \sum_{k=1}^N F_k(u,\{\mu_1 ... \mu_n\})
|\mu\rangle_{\mu_k \to u}+\sqrt{M+1}\left(\prod_{i=1}^N  \mathrm{S}^-(\mu_i) \right)\mathrm{S}^-(u) \left| M+1; \uparrow \dots \uparrow \right>$, the approach fails to simply provide the desired solution.


However, since, according to eq. (\ref{generating}), finding the common eigenstates of $ \mathrm{S}^2(u)$ $\forall \ u \in \mathbb{C}$ is formally equivalent to finding the common eigenstates of the $N$ conserved charges $R_k$, we choose to work directly with the latter. It remains simple to then go back to the spectrum of $\mathrm{S}^2(u)$ since its whole analytic structure is known. We therefore simply look at the residues (at $u\to \epsilon_k$) of the $\mathrm{S}^2(u)$ operator's application on $\prod_{i=1}^N  \mathrm{S}^-(\mu_i)\left| M;  \uparrow \dots \uparrow \right>$ to write: 
\bea
R_k|\mu\rangle  &=& \lim_{u\to \epsilon_k}(u-\epsilon_k) \mathrm{S}^2 (u)|\mu\rangle
\nonumber\\ &=&
V^2\left[\frac{1}{2}\sum_{j\neq k}\frac{1}{\epsilon_{k}-\epsilon_j}-\frac{\omega-\epsilon_k}{2V^2}-\sum_{j=1}^N\frac{1}{\epsilon_k -\mu_j}\right]|\mu\rangle \nonumber\\ && +V^2\left[\sum_{j=1}^N\frac{B_j}{\epsilon_k -\mu_j}\left(\prod_{i\ne j}^{N}  \mathrm{S}^-(\mu_i) \right)\right] S^-_k | M; \uparrow\dots \uparrow\rangle
\nonumber\\ && +V\sqrt{M+1}\left(\prod_{i=1}^N  \mathrm{S}^-(\mu_i) \right)S^-_k| M+1;  \uparrow\dots \uparrow \rangle
\label{withunwanted}
\eea

\noindent with 

\bea
B_j = \left(-\frac{\omega-\mu_j}{V}+\sum_{i=1}^N\frac{V}{\mu_j -\epsilon_i}+\sum_{i\neq j}\frac{2V}{(\mu_i -\mu_j)}\right).
\eea

Working directly with the conserved charges $R_k$ now results in a form which provides clear separation of the wanted and unwanted terms.  Indeed, the first term in (\ref{withunwanted}) will ultimately become $r_k \left|\mu_1 ... \mu_N\right>$ while the two last ones are the actual unwanted terms whose global cancellation will turn the generic ansatz into a proper eigenstate of the ensemble of $R_k$ (and therefore of $\mathrm{S}^2(u)$) with eigenvalues $r_k$.

The Gaudin lowering operator $\mathrm{S}^-(u)$, given in (\ref{sbreal}), can act either on the bosonic or on the spin subspace. Assuming for now that the number of excitations is $M \ge N-1$, one of the contributions to the unwanted terms will come from the joint action of the bosonic part of every $\mathrm{S}^-(\mu_i)$.  This reduces both kets contributing to the unwanted terms down to $S^-_k\left|M-(N-1);\uparrow\dots \uparrow\right>$. Cancelling the total coefficient in front of this particular state:

\bea
V^2\left[\sum_{j=1}^N\frac{B_j}{\epsilon_k -\mu_j}\frac{\sqrt{M!}}{\sqrt{(M+1-N)!}}\right]  +V \sqrt{M+1}\frac{\sqrt{(M+1)!}}{\sqrt{(M+1-N)!}} 
\eea

\noindent is achieved when:

\bea
\sum_{j=1}^N\frac{B_j}{\epsilon_k -\mu_j} +\frac{M+1}{V}= 0.
\eea

To form an eigenstate, the cancellation would need to occur for all possible $k \in [1,N]$ which results in a linear system of equations:

\bea
\left(
\begin{array}{cccc}
 1/(\epsilon_1-\mu_1) & 1/(\epsilon_1-\mu_2)   & \dots &1/(\epsilon_1-\mu_N)   \\
  \vdots& \vdots   &\cdots&\vdots   \\
  1/(\epsilon_N-\mu_1) & 1/(\epsilon_N-\mu_2)   & \dots  & 1/(\epsilon_N-\mu_N) 
 \end{array}
\right)
\left(
\begin{array}{c}
  B_1     \\
  \vdots   \\
  B_N     
\end{array}
\right) = 
\left(
\begin{array}{c}
  -\frac{M+1}{V}     \\
  \vdots   \\
  -\frac{M+1}{V}     
\end{array}
\right),\nonumber\\
\eea 
\noindent which is easily solved by recognizing the Cauchy matrix. Indeed, its well-known inverse leads to the solution:

\bea
B_j \equiv \frac{-\omega+\mu_{j}}{V}+\sum_{i=1}^{N}\frac{V}{\mu_{j}-\epsilon_{i}}+\sum_{i\neq j}^{N}\frac{2 V}{\mu_{i}-\mu_{j}}= -\frac{(M+1)}{V}\frac{\prod_{k=1}^N (\epsilon_{k}-\mu_{j})}{\prod_{a\neq j}^N (\mu_{a}-\mu_{j})}.
\label{bethenbosonmu}
\nonumber\\
\eea

Of course this, in principle, simply cancels one subset of coefficients. One still needs to prove that solutions to these equations will cancel every coefficient in the unwanted terms. 

A generic coefficient results from having the product of Gaudin lowering operators acting on both the bosonic and the spin part. Looking at a given ket, for which the subset of  $r +1 =N-M+m$ spins labelled $\{i_1 \dots i_{r},k\}$ points down (the corresponding number of bosons remaining being $m$), the coefficient in front of this basis vector $\left|m; \downarrow_{\{i_1 \dots i_{r},k\}}\right>$ is given by 

\bea
C^r_{\{i_1 \dots i_{r},k\}} &=& \frac{\sqrt{M!}}{\sqrt{m!}} V^{r+2}  \left(\sum_{j=1}^N\frac{B_j}{\epsilon_k -\mu_j} \left[\sum_{A_{\hat{j}} \in S^r_{\hat{j}}} \frac{1}{\prod_{k=1}^{r}(a_k - \epsilon_{i_k})}\right] \right.  \nonumber\\&& \left. \ + \frac{(M+1)}{V}\left[\sum_{A \in S^r} \frac{ 1}{\prod_{k=1}^{r}(\alpha_k - \epsilon_{i_k})} \right]\right)
\eea

\noindent where $S^r_{\hat{j}}$ is the set of all $r$-tuples built out of $r$ non-repeated elements of $\{\mu_1, \dots \mu_{j-1}, \mu_{j+1} \dots \mu_ N\}$ and $S^r$ is the similar set of all $r$-tuples  one can build out of the elements of $\{\mu_1, \dots \mu_ N\}$. The n-th element of $A_{\hat{j}}=(a_1, \dots a_r)$ is then any one of the available parameters $\mu$ (excluding $\mu_j$) while the n-th element of $A=(\alpha_1, \dots \alpha_r)$ can be any one of the $N$ parameters $\mu$ (this time including $\mu_j$).

To turn both brackets into a factor common to every term of the sum over $j$, the first one can be rewritten by extending the sum over elements of  $S^r_{\hat{j}}$ to one over every element of $S^r$. One then needs to subtract the newly introduced contributions: those which belong to the relative complement  $S^r \setminus S^r_{\hat{j}}$, i.e. $r$-tuples which DO contain $\mu_j$. The previous expression then becomes

\bea
C^r_{\{i_1 \dots i_{r},k\}} &=& \frac{V^{r+2} \sqrt{M!}}{\sqrt{m!}}  \left(\left(\sum_{j=1}^N\frac{B_j}{\epsilon_k -\mu_j} + \frac{(M+1)}{V} \right) \left[\sum_{A \in S^r} \frac{1}{\prod_{k=1}^{r}(a_k - \epsilon_{i_k})}\right] \right.  \nonumber\\&& \left.\ -\sum_{j=1}^N\frac{B_j}{\epsilon_k -\mu_j} \left[\sum_{A\in S^r \setminus S^r_{\hat{j}}} \frac{1}{\prod_{k=1}^{r}(a_k - \epsilon_{i_k})}\right] \right).\eea

The first term is cancelled by the solutions (\ref{bethenbosonmu}) discussed previously, i.e.: $B_j =  -\frac{(M+1)}{V}\frac{\prod_{k=1}^N (\epsilon_{k}-\mu_{j})}{\prod_{a\neq j}^N (\mu_{a}-\mu_{j})}$ and we are left with:

\bea
C^r_{\{i_1 \dots i_{r},k\}} \propto \sum_{j=1}^N\frac{B_j}{\epsilon_k -\mu_j} \sum_{A \in S^r \setminus S^r_{\hat{j}}} \frac{1}{\prod_{k=1}^{r}(a_k - \epsilon_{i_k})}.
\eea

Since every element $A$ in the sum now contains, with certainty, a term $a_{k'} = \mu_j$ (with $k'$ denoting the position where $\mu_j$ appears in $A$), we can rewrite the sum as:

\bea
C^r_{\{i_1 \dots i_{r},k\}} \propto \sum_{k' = 1}^r \sum_{j=1}^N\frac{B_j}{(\epsilon_k -\mu_j)(\mu_j - \epsilon_{k'})} \sum_{A \in  S^{r-1}_{\hat{j}}} \frac{1}{\prod_{k=1}^{r-1}(a_k - \epsilon_{i_k})},
\eea 
\noindent where $\epsilon_{i_k}$ is now understood as the elements of $(\epsilon_{i_1} \dots \epsilon_{i_{k'-1}} \epsilon_{i_{k'+1}}\dots  \epsilon_{i_r})$ therefore excluding $\epsilon_{i_{k'}}$.

One can keep the process going by including back $\mu_j$ in the last sum to now make it go over every element of  $S^{r-1}$ while removing $\sum_{A\in S^{r-1} \setminus S^{r-1}_{\hat{j}}}$:

\bea
C^r_{\{i_1 \dots i_{r},k\}} &\propto& \sum_{k' = 1}^r \sum_{j=1}^N\frac{B_j}{(\epsilon_k -\mu_j)(\mu_j - \epsilon_{k'})} \sum_{A \in  S^{r-1}} \frac{1}{\prod_{k=1}^{r-1}(a_k - \epsilon_{i_k})} \nonumber\\ &&- \sum_{k' = 1}^r \sum_{j=1}^N\frac{B_j}{(\epsilon_k -\mu_j)(\mu_j - \epsilon_{k'})} \sum_{A \in S^{r-1} \setminus S^{r-1}_{\hat{j}}} \frac{1}{\prod_{k=1}^{r-1}(a_k - \epsilon_{i_k})},
\nonumber\\
&=& \sum_{k' = 1}^r K_{k'} \left[\sum_{j=1}^N\frac{B_j}{(\epsilon_k -\mu_j)(\mu_j - \epsilon_{k'})} \right]\nonumber\\ &&- \sum_{k'' \ne k' }^r  \sum_{k' = 1}^r \sum_{j=1}^N\frac{B_j}{(\epsilon_k -\mu_j)(\mu_j - \epsilon_{k'})(\mu_j - \epsilon_{k''})} \sum_{A  \in S^{r-2}_{\hat{j}}} \frac{1}{\prod_{k=1}^{r-2}(a_k - \epsilon_{i_k})}\nonumber\\
\eea 

\noindent where the set of $\epsilon_{i_k}$ now excludes both $\epsilon_{i_{k'}}$ and $\epsilon_{i_{k''}}$, and we defined the constant $K_{k'} =  \sum_{A \in  S^{r-1}} \frac{1}{\prod_{k=1}^{r-1}(a_k - \epsilon_{i_k})}$. Doing this iteratively until the $r$ spins have their $\epsilon_{i_k}$ coupled with $\mu_j$, one easily sees that the whole expression becomes a sum of terms of the form:

\bea
\sum_{j=1}^N \frac{B_j}{\prod_{k=1}^{n}(\epsilon_{i_k} - \mu_j)}, 
\eea

\noindent with $n \ge 2$, which we can show are all identically zero. Indeed, replacing $B_j$ by the solution mentioned before  (\ref{bethenbosonmu}), we find:

 \bea
\sum_{j=1}^N \frac{B_j}{\prod_{k=1}^{n}(\epsilon_{i_k} - \mu_j)} = -\frac{(M+1)}{V}\sum_{j=1}^N  \frac{\prod_{k=1}^N (\epsilon_{k}-\mu_{j})}{\prod_{a\neq j}^N (\mu_{a}-\mu_{j})\prod_{k=1}^{n}(\epsilon_{i_k} - \mu_j)}
\eea

\noindent and choosing one $\epsilon_{i_{k'}}$ to play a particular role it  can also be written as:

\bea
\sum_{j=1}^N \frac{B_j}{\prod_{k=1}^{n}(\epsilon_{i_k} - \mu_j)} =-\frac{(M+1)}{V} \sum_{j=1}^N 
\frac{\frac{\prod_{k=1}^N (\epsilon_{k}-\mu_{j})}{\prod_{k\ne k'}^{n}(\epsilon_{i_k} - \mu_j)}}{\epsilon_{i_{k'}}-\mu_j} \frac{1}{\prod_{a\neq j}^N (\mu_{a}-\mu_{j})}. 
\label{lagrangezero}
\eea

Considering that a generic polynomial $L(z)$ of maximal order $N-1$ can always be expanded on the basis of Lagrange polynomials (using the $N$ points $\{\mu_1 \dots \mu_N\}$ ) as:

\bea
L(z) = \ell(z)\sum_{j=1}^N \frac{L(\mu_j)}{z-\mu_j}  \frac{1}{\prod_{k\ne j}(\mu_j-\mu_k)},
\eea

\noindent with $\ell(z) = \prod_{k=1}^N(z-\mu_k)$, we define the $(N-n+1)$-order polynomial $A(z) = \frac{\prod_{k=1}^N (\epsilon_{k}-z)}{\prod_{k\ne k'}^{n}(\epsilon_{i_k} - z)}$. Its order will be always be low enough (for $n\ge 2$) for it to have an exact expansion in the $N$-points Lagrange basis: $\displaystyle\frac{A(z)}{\ell(z)} =  \sum_{j=1}^N \frac{A(\mu_j)}{z-\mu_j}  \frac{1}{\prod_{k\ne j}(\mu_j-\mu_k)}$. We can therefore recognize in (\ref{lagrangezero}) the expression for $\frac{A(z)}{\ell(z)}$ at $z=\epsilon_{i_{k'}}$. This allows us to evaluate the preceding sum as:

\bea
\sum_{j=1}^N \frac{B_j}{\prod_{k=1}^{n}(\epsilon_{i_k} - \mu_j)} \propto \sum_{j=1}^N \frac{A(\mu_j)}{\epsilon_{i_{k'}} - \mu_j}\frac{1}{\prod_{k\ne j}(\mu_j-\mu_k)} = \frac{A(\epsilon_{i_{k'}})}{\ell(\epsilon_{i_{k'}})} = 0.
\eea

\noindent Indeed, the polynomial $A(z)$ has zeros at every $\epsilon_k$ except at the $n-1$ elements of the set $\{\epsilon_{i_1}, \dots \epsilon_{i_{k'}-1}, \epsilon_{i_{k'}+1} \dots \epsilon_{i_n}\}$. Therefore  $z=\epsilon_{i_{k'}}$ remains a zero of the polynomial $A(z)$ proving that the sum (\ref{lagrangezero}) systematically cancels when $\{\mu_1 \dots \mu_N\}$ is a solution of eqs. (\ref{bethenbosonmu}).

Thus, it has been shown that, for a system containing $N$ spins-$1/2$ and a single bosonic mode, a state built as:

\bea
\left|\{\mu_1 \dots \mu_N\}\right>= \prod_{i=1}^N \mathrm{S}^-(\mu_i) \left|M; \uparrow \dots \uparrow\right>
\eea   

\noindent becomes an eigenstate of the $N$ conserved charges $R_k$ defined in (\ref{consboson}) with corresponding eigenvalues (see eq. (\ref{withunwanted})):

\bea
r_k = V^2\left[\frac{1}{2}\sum_{j\neq k}\frac{1}{\epsilon_{k}-\epsilon_j}-\frac{\omega-\epsilon_k}{2V^2}-\sum_{j=1}^N\frac{1}{\epsilon_k -\mu_j}\right],
\label{rihole}
\eea

\noindent whenever $\{\mu_1 \dots \mu_N\}$ is solution of the $N$ algebraic Bethe equations 
(\ref{bethenbosonmu}):

\bea
\frac{-\omega+\mu_{j}}{V}+\sum_{i=1}^{N}\frac{V}{\mu_{j}-\epsilon_{i}}+\sum_{i\neq j}^{N}\frac{2 V}{\mu_{i}-\mu_{j}}= -\frac{(M+1)}{V}\frac{\prod_{k=1}^N (\epsilon_{k}-\mu_{j})}{\prod_{a\neq j}^N (\mu_{a}-\mu_{j})}.
\eea

For any of these states, the corresponding eigenvalue of the full generating function $\mathrm{S}^2(u)$ is then simply obtained by reconstructing the appropriate linear combination given in (\ref{generating}) leading to:

\bea
\mathrm{S}^2(u) \left|\{\mu_1 \dots \mu_N\}\right>=
\nonumber\\  \left[\sum_{k=1}^N \frac{r_k}{u-\epsilon_k} + M-\frac{N}{2}+ \frac{1}{2} + \left(\frac{\omega-u}{2V}\right)^2+\frac{3}{4} \sum_{k=1}^N \frac{V^2}{(u-\epsilon_k)^2}\right] \left|\{\mu_1 \dots \mu_N\}\right>,
\nonumber\\
\eea
\noindent with the eigenvalues $r_k$ given in eq. (\ref{rihole}).

Having shown that one can indeed build a valid hole-like representation of eigenstates, we now simply need to find the correspondence between both representations of a given eigenstate. As for spin-only models, this can be done by equating both sets of eigenvalues $r_k$, respectively  found in eqs. (\ref{ripart}) and (\ref{rihole}):

\bea
-\frac{\omega-\epsilon_k}{2}-V^2\sum_{j=1}^N\frac{1}{\epsilon_k -\mu_j} = 
 -\frac{\epsilon_k-\omega}{2} - V^2 \sum_{j=1}^M \frac{1}{\epsilon_k - \lambda_j}
\nonumber\\
   \Lambda^{\lambda} (\epsilon_k)
=  \Lambda^{\mu} (\epsilon_k) + \frac{\omega-\epsilon_k}{V^2}, 
\label{correspboson}
\eea

\noindent giving us the explicit way to transform one representation into the other.  A direct substitution of this transformation in the quadratic Bethe equations (\ref{quadbos}) consequently gives us the equivalent equations for the hole-like representation:

\bea \left[ \Lambda^{\mu}(\epsilon_i)\right]^2 =  \sum_{j\ne i} \frac{ \Lambda^{\mu}(\epsilon_i)- \Lambda^{\mu}(\epsilon_j)}{\epsilon_i-\epsilon_j} + \frac{\epsilon_i-\omega}{V^2}  \Lambda^{\mu}(\epsilon_i)+ \frac{M - N +1}{V^2}.
\eea

Although our main interest for finding such a representation was the construction of eigenvalue-based determinant expressions for scalar products and form factors, even by itself, the hole-representation can be useful. Indeed, such a parametrisation of the eigenstates always involves exactly $N$ rapidities $\mu_i$ which have to be solution of a system of $N$ equations. The traditional particle-representation is, on the other hand, defined by $M$ parameters $\lambda_i$, a number which can grow arbitrarily large due to the unboundedness of the bosonic operators. This alternative representation therefore still allows us to express  any eigenstate containing $M > N$ particle-like excitations in a more compact form systematically writable in terms of only $N$ complex parameters.

\section{Partition function}

As was possible for spin-only models \cite{faridet}, we want to be able to access physical quantities in terms of simple expressions involving exclusively the $\Lambda(\epsilon_i)$ variables. In this section we will start by deriving a determinant expression for domain wall boundary partition functions from which expressions for the scalar products and form factors will, in the end, be derived.

Therefore, we want to prove that the scalar product of a generic Bethe-like state (\ref{bethelikeboson}) with an arbitrary common eigenstate of $b^\dag b$ and the ensemble of $S^z_i$:
 
  \bea
  \left|M;\uparrow_{\{i_1\dots i_m\}}\right> \equiv \displaystyle (b^+)^M\prod_{j=1}^{m} S^+_{i_j} \left|0,\downarrow ... \downarrow \right>\eea
  \noindent with $M,m\in \mathbb{N}$ and $ m \leq N$, is writable as the determinant of an  $m\times m$ matrix:

\bea
\left< M;\uparrow_{\{i_1\dots i_m\}}\right.\left|\lambda_1 ... \lambda_{M+m}\right> = \sqrt{M!}V^{m}\mathrm{Det} J
\nonumber\\
\nonumber\\
J_{ab} =
 \left\{ \begin{array}{cc}
\displaystyle\sum_{c=1 (\ne a)}^m  \frac{1}{\epsilon_{i_a} -\epsilon_{i_c} }-\Lambda(\epsilon_{i_a}) & a= b
\\
\frac{1}{\epsilon_{i_a} -\epsilon_{i_b} }
& a\ne b \end{array}\right. ,
\label{detrep}
\eea

\noindent which differs significantly from the  more traditional Izergin determinant \cite{izerginold,izergin} by using the variables $\Lambda(\epsilon_{i_a})$ which are explicitly symmetric constructions in terms of the rapidities $\{\lambda_1 \dots \lambda_M\}$.

In order to show this, one can start from the explicit construction of the state $\left|\lambda_1 ... \lambda_{M+m}\right> $  (eq. (\ref{bethelikeboson})), which leads to the formal expression: 

\bea
\left< M;\uparrow_{\{i_1\dots i_m\}}\right.\left|\lambda_1 ... \lambda_{M+m}\right> = \sqrt{M!}\ V^{m}\sum_{\left\{P_k\right\}}\sum_{\left\{P\right\}} \prod_{i\neq \{k_1 \dots k_M \}}^{M+m} \frac{1}{\lambda_i-\epsilon_{P_i}}.
\label{perm}
\eea

\noindent Here $\left\{P\right\}$ is the ensemble of possible permutations of the indices $\left\{i_1 ... i_{M+m}\right\}$ and $P_i$ denotes the $i^{\mathrm{th}}$ element of the given permutation. In the same way we define $\left\{P_{k}\right\}$ as the ensemble of possible subsets $\{k_1 \dots k_M\}$ one can build out of $\{1 \dots M+m\}$; this subset labels the rapidities used to create the $M$ bosons. In other words, we create the state $\left| M;\uparrow_{\{i_1\dots i_m\}}\right>$ starting from the particle vacuum $\left| 0;\downarrow\dots \downarrow\right>$ by using any subset of $M$ rapidities to create bosons. The remaining unused rapidities can be associated in any possible bijection with $\{i_1\dots i_m\}$ so that each rapidity is used individually to excite any one of the local spins we need to flip up.

By isolating in (\ref{perm}) the terms which depend on $\lambda_{M+m}$, one finds that the overlaps obey the simple recursion relation:

\bea
\left< M;\uparrow_{\{i_1\dots i_m\}}\right.\left|\lambda_1 ... \lambda_{M+m}\right> &=& \left< M;\uparrow_{\{i_1\dots i_m\}}\right|b^\dag\left|\lambda_1 .. \lambda_{M+m-1}\right> 
\nonumber\\ & +& \sum_{j=1}^{m} \frac{V}{\lambda_{M+m} - \epsilon_{i_j}} \big< M;\uparrow_{\{i_1\dots \hat i_j \dots i_m\}}\big.\big|\lambda_1 .. \lambda_{M+m-1}\big>,\nonumber\\
\label{rec}
\eea

\noindent where $\left| M;\uparrow_{\{i_1\dots \hat i_j \dots i_m\}}\right>$ is the state with $M+m-1$ excitations, for which $\epsilon_{i_j}$ has been removed from the ensemble $\left\{i_1 \dots i_{m} \right\}$ and therefore points down. 

We now want to show that the proposed determinant representation  (\ref{detrep}) obeys the same recursion relation. Since they are obviously rational functions of every $\lambda_i$ variable, it is then sufficient to show that, when written as determinants, the left and right hand sides of the recursion (\ref{rec}) have the same poles (at $\lambda_{M+m} = \epsilon_{i_j}$), the same residues $ \big< M;\uparrow_{\{i_1\dots \hat i_j \dots i_m\}}\big.\big|\lambda_1 .. \lambda_{M+m-1}\big>$ at these poles and the same limit when $\lambda_{M+m} \to \infty$.  

 The determinant in (\ref{detrep}) clearly only has single poles at $\lambda_{M+m} = \epsilon_{i_j}$ which come only from the diagonal element $J_{jj}$ since only it  contains, through $-\Lambda(\epsilon_{i_j})$, the term $\frac{1}{\lambda_{M+m}-\epsilon_{i_j}}$. The residue is then simply given by the determinant of the minor obtained after removing line and column $j$ and taking the $\lambda_{M+m} \to \epsilon_{i_j}$ limit:

\bea
\lim_{\lambda_{M+m}\to \epsilon_{i_j}}(\lambda_{M+m} -  \epsilon_{i_j})\sqrt{M!} \ V^{m}\mathrm{Det}  J = 
\sqrt{M!}\ V^{m}\mathrm{Det}  J^{\hat{j}} 
\label{detlim}
\eea
\noindent with
\bea
J^{\hat{j}}_{a,b} = 
\left\{
\begin{array}{cc}
 \displaystyle\sum_{c=1(\ne a)}^m  \frac{1}{\epsilon_{i_a}-\epsilon_{i_c}} - \sum_{k=1}^{M+m-1} \frac{1}{\epsilon_{i_a}-\lambda_k} - \frac{1}{\epsilon_{i_a}-\epsilon_{i_j}} &  a=b \   (a,b \ne j)  \\
  \frac{1}{\epsilon_{i_a}-\epsilon_{i_b}} &  a\ne b \    (a,b \ne j)
\end{array}
\right. .\nonumber\\
\label{detpartboson}
\eea

The diagonal elements of this matrix evidently reduce to $\displaystyle\sum_{c\ne a \ (a,c \ne j) }^m \frac{1}{\epsilon_{i_a}-\epsilon_{i_c}} - \sum_{\alpha=1}^{M+m-1} \frac{1}{\epsilon_{i_a}-\lambda_\alpha}$ and therefore the limit (\ref{detlim}) does indeed correspond to the determinant representation (\ref{detrep}) of $ V \big< M;\uparrow_{\{i_1\dots \hat i_j \dots i_m\}}\big.\left|\lambda_1 .. \lambda_{M+m-1}\right> $, as it should in order to obey the recursion relation (\ref{rec}).  Since every rapidity $\lambda_j$ plays an identical role, choosing $\lambda_{M+m}$ is without loss of generality so that the determinant representations have poles and residues at those poles which are indeed the same on both sides of the recursive equation (\ref{rec}).

Moreover, the $\lambda_{M+m} \to \infty$ limit of the determinant (\ref{detrep}) is easily found since the corresponding terms $\frac{1}{\lambda_{M+m}-\epsilon_{i_a}}$ in every $\Lambda(\epsilon_{i_a})$ then simply go to zero. Therefore, $\lim_{\lambda_{M+m} \to \infty} \big< M;\uparrow_{\{i_1\dots i_m\}}\left|\lambda_1 ... \lambda_{M+m}\right> $ reduces to the determinant expression (\ref{detrep}) one would find for $\sqrt{M} \big< M-1;\uparrow_{\{i_1\dots i_m\}}\left|\lambda_1 ... \lambda_{M+m-1}\right>$.
Acting on the bra with the bosonic operator, the limit of the right hand side of (\ref{rec}) trivially gives $ \left< M;\uparrow_{\{i_1\dots i_m\}}\right| b^\dag \left|\vphantom{{\uparrow_{\{i_1\dots i_m\}}}}\lambda_1 ... \lambda_{M+m-1}\right>= \sqrt{M} \big< M-1;\uparrow_{\{i_1\dots i_m\}}\left|\lambda_1 ... \lambda_{M+m-1}\right>$ which does indeed correspond to the left hand side limit we just discussed. Consequently, if the determinant representation (\ref{detrep}) is supposed valid, the left hand side and right hand side of the recursion relation (\ref{rec}) are both rational functions of $\{\lambda_1 \dots \lambda_{M+m}\}$ and since they have the same poles, the same residues and the same limit as $\lambda_i \to \infty$, they are equal.

For $M+1$ rapidities and a single flipped spin $i_1$, the explicit expansion given in (\ref{perm}) gives the scalar product $\big<M,\uparrow_{\{i_1\}}\big.\left|\lambda_1 \dots \lambda_{M+1}\right> = \sqrt{M!} \sum_{j=1}^{M+1} \frac{V}{\lambda_j-\epsilon_{i_1}}$ which is indeed equivalent to the $1\times 1$ version of the above determinant $\sqrt{M!}\ (-\Lambda_{i_1}) = -\sqrt{M!}\sum_{j=1}^{M+1} \frac{V}{\epsilon_{i_1}-\lambda_j}$.  Since neither the order of the rapidities $\lambda_i$ nor of the $\epsilon_{i_j}$ matters, this equality concludes the recursive proof. Being valid for $M+1$ rapidities, the determinant expression can be recursively built  for $M+2$ rapidities and two flipped spins and so on, proving the validity of (\ref{detrep}).

\section{Form factors}

We showed, in the previous section, that the partition function can be written as the determinant of an $N\times N$ matrix which only depends on the eigenvalues of the conserved charges. In the current section, we now demonstrate that generic scalar products as well as form factors can also be written in terms of those partition functions, therefore giving us access to simple expressions for these quantities, again written explicitly in terms of the same variables. 

\subsection{Normalisations}

Let us first deal with the issue of the normalisation of eigenstates. As was the case for spin-only models (see section \ref{spindets}), both representations of the same eigenstate will only differ by a constant:

\bea
\left|\{\lambda_1 \dots \lambda_M\}\right> = N_\lambda \left|\{\lambda_1 \dots \lambda_M\}\right>_{\mathrm{n}}   \nonumber\\
\left|\{\mu_1 \dots \mu_N\}\right>= N_\mu \left|\{\lambda_1 \dots \lambda_M\}\right>_{\mathrm{n}} , 
\eea

\noindent  where $ \left|\{\lambda_1 \dots \lambda_M\}\right>_{\mathrm{n}} $ is the properly normalised eigenstate in question. Using the transformation (\ref{correspboson}) relating both representations, one can write the overlap (given by eq. (\ref{detrep})) as:

\bea
\left<\{\mu_1 \dots \mu_N\}\right.\left| \ \{\lambda_1 \dots \lambda_M\}\right> &=& \left<M;\uparrow \dots \uparrow \right| \prod_{i=1}^N \mathrm{S}^+(\mu_i) \prod_{j=1}^M \mathrm{S}^+(\lambda_j) \left| 0; \downarrow \dots \downarrow\right>\nonumber\\ &=& N_\lambda N_\mu = \sqrt{M!} \ V^{N}\mathrm{Det} J
\label{nbynnorm}
\eea

\noindent with

\bea
J_{ab} =
 \left\{ \begin{array}{cc}
\displaystyle \sum_{c=1 (\ne a)}^N  \frac{1}{\epsilon_{a} -\epsilon_{c} }- 2 \Lambda^\lambda(\epsilon_a)+ \frac{\omega-\epsilon_a}{V^2}& a= b
\\
\frac{1}{\epsilon_{a} -\epsilon_{b} }
& a\ne b \end{array}\right. .
\eea

\noindent Interestingly, this corresponds to the determinant of the Jacobian matrix $J_{ij} = \frac{\partial F_j}{\partial \Lambda(\epsilon_i) }$ of the set of quadratic Bethe equations (\ref{quadbos}): 

\bea F_j =\displaystyle-\left[ \Lambda(\epsilon_j)\right]^2 + \sum_{i \ne j}^M \frac{ \Lambda(\epsilon_j)- \Lambda(\epsilon_i)}{\epsilon_j-\epsilon_i}  - \frac{\epsilon_j - \omega}{V^2} \Lambda(\epsilon_j)+ \frac{M}{V^2} = 0 . 
\eea

\noindent This is, of course, highly reminiscent of the Gaudin-Korepin determinant \cite{korepinnorm} which gives the norm $\left<\lambda_1 \dots \lambda_M\right.|\left.\lambda_1 \dots \lambda_M\right>$ as the $M\times M$ determinant of the Jacobian matrix of the rapidites-based Bethe equations this time taking derivatives with respect to $\lambda_i$.

Secondly, as for spin-only models, the ratio $N_\lambda/N_\mu$ can be obtained by projecting both representations onto a reference state: a single arbitrary eigenstate common to $\{ S^z_1 \dots S^z_N\}$ and $b^\dag b$ (within the appropriate total excitation number subspace). The simplest choice could be to use $\left|M;\downarrow \dots \downarrow\right>$ as this reference state, since its scalar product with the particle-representation $\left<M;\downarrow \dots \downarrow\right.\left| \ \{\lambda_1 \dots \lambda_M\}\right> $ is trivially given by $\sqrt{M!}$; the coefficient involved coming only from the repeated action of the bosonic parts of the $\mathrm{S}^+(\lambda_i)$ operators. On the other hand, the second scalar product $\left<M;\downarrow \dots \downarrow\right.\left| \ \{\mu_1 \dots \mu_N\}\right>$ would then involve only the $N$ spin parts of every $\mathrm{S}^-(\mu_i)$ and would therefore be given by a $N\times N$ determinant.

Since the numerical calculation of the determinant of a $N\times N$ matrix requires $N^3$ operations, it would therefore be more efficient to normalise by calculating the ratio using projections on a reference state which contains  $M$ (or $N/2$ when $M>N/2$) excitations in the spin sector. This would indeed minimise the total number of operations involved in computing both determinants, which would then be given by $M^3 + (N-M)^3$ (or simply $N^3/4$ when $M>N/2$). In large scale calculations, this can become an important factor in the total computation time. 

Finally, one should keep in mind that, in certain calculations such as the resolution of the identity: $\sum_n \left|\psi_n \right>\left<\psi_n\right| = \mathbb{I}$, in which eigenstates systematically appear twice, we do not need the individual norms. Indeed, the terms can be properly normalised using only the single $N \times N$ determinant given in (\ref{nbynnorm}) by using the following constructions: $\frac{\left| \{\lambda_1 \dots \lambda_M\}\right>\left<\{\mu_1 \dots \mu_N\}\right|}{\left<\{\mu_1 \dots \mu_N\}\right.\left|  \{\lambda_1 \dots \lambda_M\}\right>}$.

\subsection{Local raising and lowering operators ($S^+_i,S^-_i,b^\dag,b$)}

The raising and lowering operators only have non-zero form factors only between two states which contain respectively $M$ and $M\pm1$ excitations. These can therefore be straightforwardly written as partition functions by using the appropriate representations of both states involved in the calculation: 

\bea
&&\left< \mu_1 \dots \mu_N \right| \mathrm{S}^+(u) \left| \lambda_1 ... \lambda_{M-1} \right> \nonumber\\ &&= \left< M ; \uparrow \dots \uparrow \right| \left(\prod_{i=1}^N \mathrm{S}^+(\mu_i)\right) \mathrm{S}^+(u)  \left(\prod_{j=1}^{M-1} \mathrm{S}^+(\lambda_j)\right) \left|0; \downarrow \dots \downarrow \right>,
\nonumber\\
&&\left<\lambda_1 ... \lambda_{M-1} \right| \mathrm{S}^-(u) \left|  \mu_1 \dots \mu_N  \right> \nonumber\\ &&= \left< 0; \downarrow \dots \downarrow \right|\left(\prod_{j=1}^{M-1} \mathrm{S}^-(\lambda_j)\right) \mathrm{S}^-(u)   \left(\prod_{i=1}^N \mathrm{S}^-(\mu_i)\right)  \left|M ; \uparrow \dots \uparrow \right>.
 \eea

These are given by the $N \times N$ determinant (\ref{detrep}) built out of the ensemble of "rapidities" $\{\nu_1 \dots \nu_{N+M}\} =\{\mu_1 \dots \mu_{N+M}, u,\lambda_1 \dots \lambda_{M-1}\}$ and give us access directly to any local spin operators by looking at the residues at $u \to \epsilon_i$ knowing that  $\lim_{u\to\epsilon_i} (u-\epsilon_k) \mathrm{S}^\pm(u) = V S^\pm_k$. Since the pole at $u \to \epsilon_k$ only appears in the diagonal element $J_{kk}$ of the partition function via the term $- \Lambda^\nu(\epsilon_k) = - \sum_{i=1}^N \frac{1}{\epsilon_k - \mu_i} - \sum_{i=1}^{M-1} \frac{1}{\epsilon_k - \lambda_i} - \frac{1}{\epsilon_k - u}$, the residue at $u = \epsilon_k$ is given by the minor determinant obtained after removing line and column $k$. Setting $u = \epsilon_k$ in the other diagonal terms lead to $- \frac{1}{\epsilon_a - u}$ cancelling the similar term in the sum 
$ \sum_{c=1 (\ne a)}^N  \frac{1}{\epsilon_{a} -\epsilon_{c}}$. We are therefore left with the simple $N-1\times N-1$ determinant:

\bea
\left< \mu_1 \dots \mu_N \right| S^+_k \left| \lambda_1 ... \lambda_{M-1} \right> =\sqrt{M!} \ V^{N-1}  \mathrm{Det} J^k ,
\eea

\noindent with

\bea
J^k_{a,b} =  \left\{ \begin{array}{cc}
\displaystyle \sum_{c=1 (\ne a,k)}^{N}  \frac{1}{\epsilon_{a} -\epsilon_{c} }-  \Lambda^\lambda(\epsilon_a)- \Lambda^\mu(\epsilon_a)& a = b
\\
\frac{1}{\epsilon_{a} -\epsilon_{b} } 
& a\ne b \end{array} \right. \ \mathrm{ with} \ (a,b \ne k).
\label{ffsplus}
\eea

The $u \to \infty$ limit gives us the bosonic operator form factors since $\lim_{u\to \infty} \mathrm{S}^+(u)  = b^\dag $. It is also easily obtained since the limit simply leads to the cancellation of the $\frac{1}{\epsilon_a - u}$ terms in each of the diagonal elements of the matrix. The resulting determinant representation is then given by the $N\times N$ determinant:

\bea
\nonumber\\
\left< \mu_1 \dots \mu_N \right| b^\dag \left| \lambda_1 ... \lambda_{M-1} \right> = \sqrt{M!} \ V^{N} \mathrm{Det} J^B,
\nonumber\\
  J^B_{a,b} =  \left\{ \begin{array}{cc}
\displaystyle \sum_{c=1 (\ne a)}^{N}  \frac{1}{\epsilon_{a} -\epsilon_{c} }-  \Lambda^\lambda(\epsilon_a)- \Lambda^\mu(\epsilon_a)& a = b
\\
\frac{1}{\epsilon_{a} -\epsilon_{b} } 
& a\ne b \end{array} \right..
\eea

\subsection{Local "number" operators $S^z_i$ and $b^\dag b$}
\label{numbersection}

The "number" operators, namely the local $S^z_i$ spin operators and the bosonic occupation operator $b^\dag b$ do not naturally allow such a simple expression. In this section, we show how one can still derive eigenvalue-based expressions for their form factors. The basic approach used here to derive such expressions is to make use of the known determinant expression (\ref{detrep}) to evaluate scalar products between an eigenstate $\left|\{\mu(\omega)\}_m\right>$ of the system for a given parameter $\omega$ and a second one $\left|\{\lambda(\omega+d\omega)\}_n\right>$ evaluated at $\omega+ d\omega$. This second eigenstate then corresponds to an infinitesimal deformation of the eigenstate of interest: $\left|\{\lambda(\omega)\}_n\right>$ found at $\omega$.  

Starting from the explicit expression of the conserved charges of the model (eq. (\ref{consboson})), we have $S^z_k = -\frac{\partial R_k(\omega)}{\partial \omega} = \frac{-R_k(\omega+d\omega) + R_k(\omega)}{d\omega}$ and therefore can deduce the spin form factors from the evaluation of:

\bea
 \left<\{\mu(\omega)\}_m\right|S^z_k\left|\{\lambda(\omega+d\omega)\}_n\right> 
 &=& \left[r_k^m(\omega)-r_k^n(\omega+d\omega) \right]\frac{\left<\{\mu(\omega)\}_m\right.\left|\{\lambda(\omega+d\omega)\}_n\right>}{d\omega}.
 \nonumber\\
\eea

Using the explicit eigenvalues (\ref{ripart}) and (\ref{rihole}), the diagonal form factors which involve the same state in the bra and the ket will then lead to the typical Hellmann-Feynman theorem:

\bea
\left<\{\mu(\omega)\}_m\right|S^z_k\left|\{\lambda(\omega)\}_m\right> &=& -\frac{\partial r^m_k(\omega)}{\partial \omega} \left<\{\mu(\omega)\}_m\right.\left|\ \{\lambda(\omega)\}_m\right>,
\nonumber\\ 
\frac{\left<\{\mu(\omega)\}_m\right|S^z_k\left|\{\lambda(\omega)\}_m\right>}{\left<\{\mu(\omega)\}_m\right.\left|\ \{\lambda(\omega)\}_m\right>}&=&\left[ -\frac{1}{2} + \frac{\partial \Lambda^\lambda_m(\epsilon_k)}{\partial \omega}\right] .
\eea

\noindent while the off-diagonal ones give:

\bea
 &&\left<\{\mu(\omega)\}_m\right|S^z_k\left|\{\lambda(\omega)\}_n\right> = \left[r_k^m(\omega)-r_k^n(\omega) \right]\frac{\left<\{\mu(\omega)\}_m\right.\left|\{\lambda(\omega+d\omega)\}_n\right>}{d\omega}
\nonumber\\
&&=
\left[\epsilon_k-\omega +V^2\Lambda^\lambda_n(\epsilon_k)- V^2\Lambda^\mu_m(\epsilon_k)\right]\frac{\left<\{\mu(\omega)\}_m\right.\left|\{\lambda(\omega+d\omega)\}_n\right>}{d\omega},\  m\ne n. \nonumber\\
\eea

One notices that a single evaluation of $\frac{\left<\{\mu(\omega)\}_m\right.\left|\{\lambda(\omega+d\omega)\}_n\right>}{d\omega}$ gives direct access to every one of the $N$ local $S^z_i$ form factors. Since the Gaudin raising operators $\mathrm{S}^+(u)$ does not explicitly depend on the parameter $\omega$, the overlap of interest is given once more by the same partition function whose determinant expression (\ref{detrep}) has been given before. The deformation to $\omega+d\omega$ indeed only appears indirectly through the infinitesimal modification of the $\lambda$ rapidities. This scalar product could, in principle, be numerically evaluated as a single determinant using the eigenvalues defining exactly the states respectively at $\omega$ and $\omega+\Delta \omega$ (with $\Delta \omega$ being small but finite):

\bea
\lim_{\Delta \omega \to 0}\frac{\left<\{\mu(\omega)\}_m\right.\left|\{\lambda(\omega+\Delta \omega)\}_n\right>}{\Delta \omega} =  \sqrt{M!} \ V^{N} \frac{1}{\Delta \omega}\lim_{\Delta \omega \to 0} \mathrm{Det} J
\eea

\noindent with:

\bea
  J_{a,b} =  \left\{ \begin{array}{cc}
\displaystyle \sum_{c=1 (\ne a)}^{N}  \frac{1}{\epsilon_{a} -\epsilon_{c} }-  \left.\Lambda_n^\lambda(\epsilon_a)\right|_{\omega+\Delta\omega}- \left.\Lambda_m^\mu(\epsilon_a)\right|_{\omega}& a = b
\\
\frac{1}{\epsilon_{a} -\epsilon_{b} } 
& a\ne b \end{array} \right..
\eea

Being of order $\Delta \omega$, the resulting determinant would however then be nearly singular, a fact which could lead to important numerical stability issues. However, an exact evaluation remains possible by taking the limit analytically using $\left.\Lambda_n^\lambda(\epsilon_a)\right|_{\omega+\Delta\omega} \approx \left.\Lambda_n^\lambda(\epsilon_a)\right|_{\omega}+\frac{\partial\Lambda_n^\lambda(\epsilon_a) }{\partial \omega}\Delta \omega$ and the fact that $\left<\psi_m(\omega)\right.\left|\psi_n(\omega)\right> = 0$ since both eigenstates are distinct and therefore orthogonal. Retaining the linear terms leads to the following sum of $N$ determinants:

\bea
\lim_{\Delta \omega \to 0}\frac{\left<\{\mu(\omega)\}_m\right.\left|\{\lambda(\omega+\Delta \omega)\}_n\right>}{\Delta \omega}  = \sqrt{M!}\ V^{N} \sum_{k=1}^N \frac{\partial \Lambda_n^\lambda(\epsilon_k) }{\partial \omega}\mathrm{det} \tilde{J}^k,
\label{sumofdets}
\eea

\noindent which are just the $N-1 \times N-1$ minors obtained from $J$ after removing line and column $k$:

\bea
\tilde{J}^k_{ab} =   \left\{ \begin{array}{cc}
\displaystyle \sum_{c=1 (\ne a)}^{N}  \frac{1}{\epsilon_{a} -\epsilon_{c} }-  \Lambda^\lambda(\epsilon_a)- \Lambda^\mu(\epsilon_a)& a = b
\\
\frac{1}{\epsilon_{a} -\epsilon_{b} } 
& a\ne b \end{array} \right.   \ \mathrm{ with} \ (a,b \ne k).
\eea

Notice that they differ from the form factors for $S^+_k$ found in eq. (\ref{ffsplus}) by the fact that the sum over $\epsilon_c$ still includes the term $c=k$. The expression found here depends not only on the eigenvalues  $\Lambda^\lambda(\epsilon_i)$ but also on their derivatives $\frac{\partial \Lambda^\lambda(\epsilon_i)}{\partial \omega}$. Fortunately, they  can be directly obtained from the knowledge of the set $\{\Lambda^\lambda(\epsilon_1) \dots \Lambda^\lambda(\epsilon_N)  \}$ by simply solving the linear system of equations:

\bea
2  \Lambda^\lambda(\epsilon_i)\frac{\partial \Lambda^\lambda(\epsilon_i)}{\partial \omega} = \sum_{j \ne i} \frac{ \frac{\partial \Lambda^\lambda(\epsilon_i)}{\partial \omega}- \frac{\partial \Lambda^\lambda(\epsilon_j)}{\partial \omega}}{\epsilon_i-\epsilon_j}  -\frac{\epsilon_i - \omega}{V^2} \frac{\partial \Lambda^\lambda(\epsilon_i)}{\partial \omega} +\frac{\Lambda^\lambda(\epsilon_i)}{V^2} 
\label{derilin}
\eea

\noindent which one finds by taking the $\omega$ derivative of the quadratic Bethe equations (\ref{quadbos}).

From the fact that $b^\dag b + \sum_{i=1}^N S^z_i$ is also a conserved quantity, with eigenvalue $M-\frac{N}{2}$ for states containing $M$ particle-like excitations, one can directly find a similar expression for the bosonic occupation form factor:
\bea
\left<\{\mu(\omega)\}_m\right|b^\dag b\left|\{\lambda(\omega\}_n\right> &=& \left<\{\mu(\omega)\}_m\right| M-\frac{N}{2}\left|\{\lambda(\omega\}_n\right> \nonumber\\ && - \sum_{i=1}^N \left<\{\mu(\omega)\}_m\right| S^z_i\left|\{\lambda(\omega\}_n\right>.
\eea

 \noindent Using the previous $S^z$ form factors we have
\bea
\frac{\left<\{\mu(\omega)\}_m\right|b^\dag b\left|\{\lambda(\omega\}_m\right> }{\left<\{\mu(\omega)\}_m\right.\left|\ \{\lambda(\omega)\}_m\right>} &=&M-\frac{N}{2} + \sum_{i=1}^{N}\frac{\partial r^m_i(\omega)}{\partial \omega}=M- \sum_{i=1}^N\left[ \frac{\partial \Lambda^\lambda_m(\epsilon_i)}{\partial \omega}\right],\nonumber\\
\eea
\bea
\left<\{\mu(\omega)\}_m\right|b^\dag b\left|\{\lambda(\omega\}_n\right> &=& \left(\sum_{i=1}^N \left[r_i^n(\omega)-r_i^m(\omega) \right]\right)\nonumber\\ && \times \frac{\left<\{\mu(\omega)\}_m\right.\left|\{\lambda(\omega+d\omega)\}_n\right>}{d\omega}, \ \ \  m \ne n\eea

\noindent which is once again proportional to the term $\frac{\left<\{\mu(\omega)\}_m\right.\left|\{\lambda(\omega+d\omega)\}_n\right>}{d\omega}$ and can therefore be explicitly computed as eq. (\ref{sumofdets}).

Let us finally mention that an identical construction allows one to build similar expressions for the $S^z_i$ form factors of spin-only realisations. By using the $V = \frac{1}{g}$ derivatives of their conserved charges given in eq. (\ref{conservedspin}), we have  $S^z_i = -\frac{1}{2}\frac{\partial R_k}{\partial V}$. Using the overlaps between eigenstates at $V$ and $V+dV$, one can then write the corresponding form factors either as a $\Delta V \to 0$ limit or explicitly as a sum of $N$ determinant minors multiplied by $\frac{\partial \Lambda^\lambda(\epsilon_i)}{\partial V}$. These derivatives can again be found by solving a simple linear system of equations obtained by deriving the relevant quadratic Bethe equations with respect to $V$.

\section{Conclusions}

In this work it was shown that, despite the lack of symmetry between the "highest" and "lowest" weight state of quantum integrable models derived from a realisation of the rational generalized Gaudin algebra which contains a bosonic mode, one can still use the QISM  to build two distinct algebraic Bethe ans\"{a}tze and therefore two representations of eigenstates of DJCG-models. These were then used to relate scalar products of states and form factors of local operators to a domain wall boundary partition function which was shown to have a simple expression as the determinant of a matrix whose elements depend only on the eigenvalues of the model's conserved charges. 

The determinant expressions derived within this paper can have an important positive impact on the computation time required for numerical work on these systems. Indeed, being defined only through these eigenvalues which are solutions of a set of quadratic Bethe equations, they allow one to avoid the explicit finding of the rapidities describing a given state $\{\lambda_1 \dots \lambda_M\}$. This work further establishes our capacity to rebuild many aspects of the algebraic Bethe ansatz using only constructions which are explicitly symmetric in the Bethe rapidities, a useful fact which could possibly generalise to a much broader class of integrable systems.

\end{document}